# Making an analogy between a multi-chain interaction in Charge Density Wave transport and the use of wave functionals to form S-S' pairs

A. W. Beckwith
Department of Physics and Texas Center for Superconductivity and Advanced Materials at the University of Houston
Houston, Texas 77204-5005, USA

## ABSTRACT

First, we show through a numerical simulation that the massive Schwinger model used to formulate solutions to CDW transport in itself is insufficient for transport of soliton-antisoliton (S-S') pairs through a pinning gap model of CDW transport. We show that a model Hamiltonian with Peierls condensation energy used to couple adjacent chains (or transverse wave vectors) permits formation of S-S' pairs which could be used to transport CDW through a potential barrier .Previously, we have argued that there are analogies between this construction and the false vacuum hypothesis used for showing a necessary and sufficient condition for formation of CDW S-S' pairs in wavefunctionals. Here we note that this can be established via either use of the Bogomil'nyi inequality or an experimental artifact which is due to use of .the false vacuum hypothesis to obtain a proportional 'distance' between the S-S' charge centers.

# I. INTRODUCTION

We have prior to this paper formed an argument using the integral Bogomol'nyi inequality to present how a soliton-anti soliton (S-S') pair could form.[1,2] In addition, we also have shown how the formation of wave functionals is congruent with Lin's nucleation of an electron-positron pair as a sufficiency argument as to forming Gaussian wave functionals. Here, we argue our wavefunctional result is equivalent to putting in a multi-chain interaction term in our simulated Hamiltonian system. with a constant term in it proportional to the Peierls gap times a cosine term representing interaction of different CDW chains in our massive Schwinger[3] model. This change of the Hamiltonian term is adding in an additional potential energy term making the problem look like a Josephon junction problem. We found that a single-chain simulation of the S-S' transport problem suffers from two defects. First, it does not answer what are necessary and sufficient conditions for formation of a. S-S' pair. More importantly, we also find through numerical simulations of the single-chain transport model that one needs additional physical conditions to permit barrier penetration. Our numerical simulation of the single-chain problem for CDW involving S-S' pairs gave a resonance condition in transport behavior over time, with no barrier tunneling. The argument here that we will present is that the false vacuum hypothesis[1,2,4] is a necessary condition for the formation of S-S' pairs and that the multi-chain term we add to a massive Schwinger equation for CDW transport is a sufficiency condition for the explicit formation of a soliton (anti soliton) in our charge density wave transport problem. We begin this by a numerical simulation of the single-chain model of CDW, then show how addition of the Peierls condensensation energy permits a soliton (anti soliton) to form .We finally discuss in the last part of the paper how this would tie in with either the Bogomil'nyi inequality and/or the phenomenological Gaussian wave functional model of S-S' pair formation and would permit necessary additional conditions to permit CDW dynamics approaching what we see in the laboratory Appendix I below gives a summary of how to computationally simulate multi-chains, while the general argument ties the analysis of this problem field theoretically to methods presented in my dissertation and in other articles under editing review.

## II. REVIEW OF THE NUMERICAL BEHAVIOR OF A SINGLE-CHAIN FOR CDW DYNAMICS

We are modifying a one chain model of Charge Density Wave (CDW) transport initially pioneered by Dr. John Miller[5] which furthered Dr. John Bardeens work[6] on a pinning gap presentation of CDW transport. The single-chain model is a good way to introduce how a threshold electric field would initiate transport, qualitatively speaking. We did, however, when using it, assume that the CDW would be easily modeled with a soliton (anti soliton) Gaussian packet. So we undertook this investigation to determine necessary and sufficient condition to physically justify use of a soliton (anti-soliton) for our wave packet. We start by using an extended Schwinger model [3] with the Hamiltonian set as

$$H = \int_x \left[ \frac{1}{2 \cdot D} \cdot \Pi_x^2 + \frac{1}{2} \cdot (\partial_x \phi_x)^2 + \frac{1}{2} \cdot \mu_E^2 \cdot (\phi_x - \varphi)^2 + \frac{1}{2} \cdot D \cdot \omega_P^2 \cdot (1 - \cos\phi) \right] \quad (2.1)$$

We should note in writing this that that a washboard potential with small driving term[6] $\mu_E \cdot (\phi - \Theta)^2$ added to the main potential term of the washboard potential, is used to model transport phenomenology. We also argue that this potential permits domain wall modeling of S-S' pairs.[7] In this situation, $\mu_E$ is proportional to the electrostatic energy between the S-S' pair constituents (assuming a parallel plate capacitor analogy) ; $\epsilon$ is a small driving force we will explain later, dependent upon a ratio of an applied electric field over a threshold field value. As we show later, the dominant washboard potential term will have the value of (pinning energy) times $(1-\cos\phi)$. We call the $V_E$ the Euclidian action version of the potential given above. In addition, the first term in this Eq. (2.1) is the conjugate momentum. Specifically, we found that we had $\Pi_x \equiv D \cdot \partial_t \phi_x$ as canonical momentum density, $D \equiv \left( \dfrac{\mu \cdot h}{4 \cdot \pi \cdot v_F} \right)$, (where $\mu \equiv \dfrac{M_F}{m_{e^-}} \cong 10^3$ is a Frohlich to electron mass ratio, and $v_F$ is a Fermi velocity $> 10^3$ cm/sec), and $D \cdot \omega_P^2$ as the pinning energy. In addition, we have that $\mu_E$ is electrostatic energy, which is analogous to having a S-S' pair represented by a separation L and of cross-sectional area A, which produces an internal field $E^* = (e^*/\varepsilon \cdot A)$, where $e^\bullet \cong 2 \cdot e^- \equiv$ effective charge and $\varepsilon \equiv 10^8 \cdot \varepsilon_0$ is a huge dielectric constant. Finally, the driving force term, $\Theta = 2 \cdot \pi \cdot \dfrac{E}{E^*}$, where the physics of the term given by $\int dx \cdot \mu_E \cdot (\phi - \Theta)^2$, leads to no instanton tunneling transitions if $\Theta < \pi \Leftrightarrow E < \dfrac{E^*}{2}$ which was the basis of a threshold field of the value $E_T = E^*/2$ due to conservation of energy considerations. Finally, it is important to note that experimental constraints as noted in the device development laboratory lead to $.01 < \mu_E/D \cdot \omega_P^2 \leq .015$, which we claim has also been shown to be necessary due to topological soliton arguments.

It is useful to note that Kazumi Maki,[8] in 1977, gave the first generalization of Sidney Coleman's[9] least action arguments to NbSe$_3$ electrodynamics. We use much the same pinning potential, with an additional term due to capacitance approximation of energy added by the interaction of a S-S' pair with each other.[5,6] While Dr. Maki's work is very complete, it does not include in a feature we found of paramount importance, that of the effects of a threshold electric field value to 'turn on' effective initiation of S-S' pair transport across a pinning gap. We should note the new physics added here since in this situation, $\mu_E$ is proportional to the electrostatic energy between the S-S' pair constituents (assuming a parallel plate capacitor analogy); $\in$ is a small driving force we will explain later, dependent upon a ratio of an applied electric field over a threshold field value. It is also relevant to note that we previously found[10] that topological soliton style arguments can explain why the potential lead to the least action integrand collapsed to primarily a quadratic potential contribution, which permits treating the wave functional as a Gaussian .As would be expected ,the ratio of the coefficient of pinning gap energy of the Washboard potential used in NbSe$_3$ modeling to the quadratic term $\mu_E \cdot (\phi - \Theta)^2$ used in modeling energy stored in between S-S' pairs was fixed by experiment to be nearly 100 to 1 , which is a datum we used in our calculations.[11]

To those whom are unfamiliar with the Schwinger model, we can summarize it briefly as follows. Namely, we use that the Schwinger model, named after Julian

Schwinger, is the model describing 2D Euclidean quantum electrodynamics with a Dirac fermion. This model exhibits a spontaneous symmetry breaking of the U(1) symmetry due to a chiral condensate due to a pool of instantons. The photon now becomes a massive particle. This model can be solved exactly and is used as a toy model for other more complex theories. We use it, keeping in mind the instanton[12] flavor to the model, as well as how instantons can be analytically conveyed in transport via a wave functional with a Gaussian integrand[13] and work with a quantum mechanically based energy

$$E = i\hbar \frac{\partial}{\partial t} \quad (2.2a)$$

and momentum

$$\Pi = (\hbar/i) \cdot \partial/\partial \phi(x) \quad (2.2b)$$

The first case is a one-chain mode situation. Here, $\epsilon \equiv a_D t$ was used explicitly as a driving force, while using the following difference equation due to using the Crank Nickelson[14] scheme. We should note that $a_D$ is a driving frequency to this physical system which we were free to experiment with in our simulations. The first index, j, is with regards to 'space', and the second, n, is with regards to 'time' step. Eq. (2.3) is a numerical rendition of the massive Schwinger model plus an interaction term, where one is calling $E = i\hbar \frac{\partial}{\partial t}$ and one is using the following replacement

$$\begin{aligned}\phi(j,n+1) = &\phi(j,n-1) \\ &+ i\cdot\Delta t\cdot\left(\frac{\hbar}{D}\left[\frac{\phi(j+1,n)-\phi(j-1,n)-2\cdot\phi(j,n)+\phi(j+1,n+1)+\phi(j-1,n+1)-2\phi(j,n+1)}{(\Delta x)^2}\right]\right. \\ &\left. -\frac{2\cdot V(j,n)}{\hbar}\phi(j,n)\right)\end{aligned} \quad (2.3)$$

We use variants of Runge-Kutta[14] in order to obtain a sufficiently large time step interval so as to be able to finish calculations in a reasonable period of time; this avoids an observed spectacular blow up of simulated average phase values; which was observed after 100 time steps at $\Delta t \approx 10^{-13}$. Stable Runge-Kutta simulations require $\Delta t \approx 10^{-19}$: A second numerical scheme, the Dunford-Frankel[14] and 'fully implicit allows us to expand the time step even further. Then, the 'massive Schwinger model' is:

$$\phi(j,n+1) = \frac{2\cdot\widetilde{R}}{1+2\cdot\widetilde{R}}\cdot(\phi(j-1,n)-\phi(j+1,n)) + \frac{1-2\cdot\widetilde{R}}{1+2\cdot\widetilde{R}}\cdot\phi(j,n-1) \quad (2.4)$$

$$-i\cdot\Delta t\frac{V(j,n)}{\hbar}\phi(j,n)$$

where $\widetilde{R} = -i\cdot\Delta t\frac{\hbar}{2\cdot D\cdot(\Delta x)^2}$. The advantage of this model is that it is second order accurate, explicit, and unconditionally stable, so as to avoid numerical blow up behavior. One then gets resonance phenomena as represented by Fig. 1. This is quite unphysical and necessitates making changes, which we will be presenting in this manuscript. In particular, we observed Eq. (2.4) results in a run away oscillation which corresponds to a continual adding up of non dissipated energy of a S-S' pair bouncing between the walls of the potential system, without tunneling commencing. Appendix I refers to the run away resonance phenomena effect for one chain and also describes how numerical simulations for more than one chain can be organized, with results finalized in Fig. 9 of the manuscript below. Let us now refer to analytical derivations needed to alter the numerical short comings of the single-chain charge density wave model

## III. ADDITION OF AN NEW TERM IN THE MASSIVE SCHWINGER EQUATION TO PERMIT FORMATION OF A S-S' PAIR

Initially we will present how addition of an interaction term between adjacent CDW chains will allow a soliton (anti soliton) to form due to some analytical considerations we will present here.[11] Finally we shall endeavor to show how our argument with the interaction term ties in with the fate of the false vacuum construction of S-S' terms done in our prior publication where either one used the Bogomil'nyi inequality[15] as a necessary condition to the formation of S-S' terms or used the ground state ansatz argument which still uses the false vacuum hypothesis extensively. Let us now first refer to how we can obtain a soliton via assuming that adjacent CDW terms can interact with each other.

One of our references,[1,2] uses the Bogomil'nyi inequality to obtain a S-S' pair which we approximate via a thin wall approximation and the nearest neighbor approximation of how neighboring chains interrelate with one another to obtain a representation of phase evolution as an arctan function w.r.t. space and time variables. Another uses the equivalence of the false vacuum hypothesis with the existence of ground state wave functionals in a Gaussian configuration.[10] To whit, either the false vacuum hypothesis itself creates conditions for the necessity of a Gaussian ansatz, or else the Bogomil'nyi inequality provides for the necessity of a S-S' pair nucleating via a Gaussian approximation which is the only way to answer data Dr. Miller collected in an experiment in 1985.[16]. But in our separate model presented in this paper we find that the interaction of neighboring chains of CDW material permits the existence of solitons (anti-solitons) in CDW transport due to the huge $\Delta'$ term added which lends to a Josephon junction interpretation of this transport problem in CDW dynamics.

Note that in the argument about the formation of a soliton (anti soliton), that we use a multi-chain simulation Hamiltonian with Peierls condensation energy used to couple adjacent chains (or transverse wave vectors) as represented by

$$H = \sum_n \left[ \frac{\Pi_n^2}{2 \cdot D_1} + E_1[1 - \cos\phi_n] + E_2(\phi_n - \Theta)^2 + \Delta' \cdot [1 - \cos(\phi_n - \phi_{n-1})] \right] \qquad (3.1a)$$

with 'momentum' we define as

$$\Pi_n = (\hbar/i) \cdot \partial/\partial\phi_n \qquad (3.1b)$$

We then use a nearest neighbor approximation to use a Lagrangian based calculation of a chain of pendulums coupled by harmonic forces to obtain a differential equation which has a soliton solution. To do this, we write the interaction term in the potential of this problem as

$$\Delta'(1 - \cos[\phi_n - \phi_{n-1}]) \to \frac{\Delta'}{2} \cdot [\phi_n - \phi_{n-1}]^2 + \text{very small H.O.T.s.} \qquad (3.2)$$

and then consider a nearest neighbor interaction behavior via

$$V_{n.n.}(\phi) \approx E_1[1 - \cos\phi_n] + E_2(\phi_n - \Theta)^2 + \frac{\Delta'}{2} \cdot (\phi_n - \phi_{n-1})^2 \qquad (3.3)$$

Here, we set $\Delta' \gg E_1 \gg E_2$, so then

$$V_{n.n.}(\phi)\bigg|_{\substack{first \\ order \\ roundoff}} \approx E_1[1-\cos\phi_n] + \frac{\Delta'}{2}\cdot(\phi_{n+1}-\phi_n)^2 \qquad (3.4)$$

which then permits us to write

$$U \approx E_1 \cdot \sum_{l=0}^{n+1}[1-\cos\phi_l] + \frac{\Delta'}{2}\cdot\sum_{l=0}^{n}(\phi_{l+1}-\phi_l)^2 \qquad (3.5)$$

which allowed using $L = T - U$ a Lagrangian based differential equation of

$$\ddot{\phi}_i - \omega_0^2[(\phi_{i+1}-\phi_i)-(\phi_i-\phi_{i-1})] + \omega_1^2 \sin\phi_i = 0 \qquad (3.6)$$

with

$$\omega_0^2 = \frac{\Delta'}{m_{e^-}l^2} \qquad (3.7)$$

and

$$\omega_1^2 = \frac{E_1}{m_{e^-}l^2} \qquad (3.8)$$

where we assume the chain of pendulums, each of length $l$, leads to a kinetic energy

$$T = \frac{1}{2}\cdot m_{e^-}l^2 \cdot \sum_{j=0}^{n+1}\dot{\phi}_j^2 \qquad (3.9)$$

where we neglect the E₂ value. However, having $E_2 \to \varepsilon^+ \approx 0^+$ would tend to lengthen the distance between a S-S' pair nucleating, with a tiny value of $E_2 \to \varepsilon^+ \approx 0^+$ indicating that the distance L between constituents of a S-S' pair would get very large

We did find that it was necessary to have a large $\Delta'$ for helping us obtain a Sine-Gordon equation. This is so if we set the horizontal distance of the pendulums to $d$, then we have that the chain is of length $L' = (n+1)d$. Then, if mass density is $\rho = m_{e^-}/d$ and we model this problem as a chain of pendulums coupled by harmonic forces, we set an imaginary bar with a quantity $\eta$ as being the modulus of torsion of the imaginary bar, and $\Delta' = \eta/d$. We have an invariant quantity, which we will designate as: $\omega_0^2 d^2 = \dfrac{\eta}{\rho \cdot l^2} = v^2$, which, as n approaches infinity, allows us to write a Sine-Gordon equation

$$\frac{\partial^2 \phi(x,t)}{\partial t^2} - v^2 \frac{\partial \phi(x,t)}{\partial x^2} + \omega_1^2 \sin \phi(x,t) = 0 \tag{3.10}$$

with a way to obtain soliton solutions. We introduce dimensionless variables of the form $z = \dfrac{\omega_1}{v} \cdot x$, $\tau = \omega_1 \cdot t$, leading to a dimensionless Sine–Gordon equation we write as

$$\frac{\partial^2 \phi(z,\tau)}{\partial \tau^2} - \frac{\partial^2 \phi(z,\tau)}{\partial z^2} + \sin \phi(z,\tau) = 0 \tag{3.11}$$

so that

$$\phi_\pm(z,\tau) = 4 \cdot \arctan\left(\exp\left\{\pm \frac{z + \beta \cdot \tau}{\sqrt{1-\beta^2}}\right\}\right) \tag{3.12}$$

where the value of $\phi_\pm(z,\tau)$ is between 0 to $2 \cdot \pi$. As an example of how we can do this value setting, consider if we look at $\phi_+(z,\tau)$ and set $\beta = -.5$. If $\tau = 0$ we can have $\phi_+(z \ll 0, \tau = 0) \approx \varepsilon \approx 0$ and also have $\phi_+(z = 0, \tau = 0) = \pi$, whereas for sufficiently large $z$ we can have $\phi_+(z, \tau = 0) - 2 \cdot \pi$. In a diagram with z as the abscissa and $\phi_4(z,\tau)$ as the ordinate, this propagation of this soliton 'field' from 0 to $2 \cdot \pi$ propagates with increasing time in the positive z direction and with a dimensionless 'velocity' of $\beta$. In terms of the original variables, one has that the 'soliton' so modeled moves with velocity $v \cdot \beta$ in either the positive or negative x direction. One gets a linkage with the original pendulum model linked together by harmonic forces by allowing the pendulum chain as an infinitely long rubber belt whose width is $l$ and which is suspended vertically. What we have described is a flip over of a vertical strip of the belt from $\phi = 0$ to $\phi = 2 \cdot \pi$ which moves with a constant velocity along the rubber belt.. First, we are using the nearest neighbor approximation to simplify Eq. (3.4) .Then, we are assuming that the contribution to the potential due to the driving force $E_2(\phi_n - \Theta)^2$ is a second order effect. All of this makes for the 'capacitance' effect given by $E_2(\phi_n - \Theta)^2$ not being a decisive influence in deforming the solution, and is a second order effect. This 2$^{nd}$ order effect contribution is enough to influence the energy band structure the soliton will be tunneling through but is not enough to break up the soliton itself. We can see how this fits into density wave transport by looking at Fig. 2 which gives us a good summary of how density waves transport themselves through a solid. We will in the next section develop a discussion about this while using a momentum space representation of a soliton- anti soliton pair(S-S') using a momentum space representation of soliton- anti soliton pair(S-S'), i.e. via a Fourier transform in momentum space of a phase we call in position space

$$\phi(x) = \pi \cdot [\tanh b(x - x_a) + \tanh b(x_b - x)] \quad (3.13)$$

# IV. WAVE FUNCTIONAL PROCEDURE USED IN S-S' PAIR NUCLEATION

Traditional current treatments frequently follow the Fermi golden rule for current density

$$J \propto W_{LR} = \frac{2 \cdot \pi}{\hbar} \cdot |T_{LR}|^2 \cdot \rho_R(E_R) \tag{4.1}$$

In our prior work we applied either the Bogomil'nyi inequality [1,2,3,,6] or we did more heuristic procedures with Gaussian wave functionals as Gaussian ansatz's to come up with an acceptable wave functional, which will refine I-E curves[2,3] used in density wave transport. For the Bogomol'nyi inequality approach we modify a de facto 1+1 dimensional problem in condensed matter physics to being one which is quasi one dimensional by making the following substitution, namely looking at the Lagrangian density $\varsigma$ to having a time independent behavior denoted by a sudden pop up of a S-S' pair via the substitution of the nucleation 'pop up' time by

$$\int d\tau \cdot dx \cdot \varsigma \to t_P \cdot \int dx \cdot L \tag{4.2}$$

where $t_P$ is the Planck's time interval. Then afterwards, we shall use the substitution of $\hbar \equiv c \equiv 1$ so we can write

$$\psi \propto c \cdot \exp\left(-\beta \cdot \int L \, dx\right) \tag{4.3}$$

This was later generalized to be of the form in a momentum space DFT momentum basis in an initial physical state with

$$\alpha \cdot \int dx [\phi_0 - \phi_C]^2_{\phi_C \equiv \phi_T} \equiv \left(\frac{2 \cdot \pi}{L}\right)^2 \cdot \sum_n |\phi(k_n)|^2 \tag{4.4a}$$

and a DFT representation of

$$\alpha \cdot \int dx [\phi_0 - \phi_C]^2_{\phi_C \equiv \phi_F} \equiv \left(\frac{2 \cdot \pi}{L}\right)^2 \cdot \sum_n (1 - n_1^2) \cdot |\phi(k_n)|^2 \tag{4.4b}$$

These in the Charge Density wave case assumed later on that $\phi(k)$ was a momentum space Fourier transform of a soliton-anti soliton pair(S-S') and that $n_1 \approx 1 - \varepsilon^+ < 1$ represented the height of this pair reaching its nucleation value, while $a \approx L^{-1}$ was one over the distance between positive and negative charge centers of the S-S' pair. Furthermore, in our case we found that in the general Gaussian wave functional ansatz approach, best to assume this, more or less, is a ground state energy start to a one dimensional Hamiltonian of a character which will lead to analytical work in momentum space leading to functional current we derived as being of the form[17]

$$J \propto T_{if} \tag{4.5}$$

This actually became a modulus argument due to considering a current density proportional to $|T|$ rather than $|T|^2$ since tunneling, in this case, would involve coherent transfer of individual (first-order) bosons rather than pairs of fermions. We used functional integral methods to extend this, in momentum space to obtain the final expression which was used after we changed the Hamiltonian tunneling element to become[3,18]

$$T_{if} \cong \frac{(\hbar^2 \equiv 1)}{2 \cdot m_e} \int \left( \Psi^*_{initial} \frac{\delta^2 \Psi_{final}}{\delta \phi(x)_2} - \Psi_{final} \frac{\delta^2 \Psi^*_{initial}}{\delta \phi(x)_2} \right) \vartheta(\phi(x) - \phi_0(x)) \wp \phi(x) \tag{4.6}$$

where we are interpreting $\wp \phi(x)$ to represent taking integration over a variation of paths in the manner of quantum field theory, and $\vartheta(\phi(x) - \phi_0(x))$ is a step function indicating that we are analyzing how a phase $\phi(x)$ evolves in a pinning gap style potential barrier. We are assuming quantum fluctuations about the optimum configurations of the field $\phi_F$ and $\phi_T$, while $\phi_0(x)$ represents an intermediate field configuration inside the tunnel barrier as we represented by Fig 3. We pick in both approaches wave functionals with

$$c_2 \cdot \exp\left(-\alpha_2 \cdot \int d\tilde{x} [\phi_T]^2 \right) \cong \Psi_{final} \tag{4.7}$$

and

$$c_1 \cdot \exp\left(-\alpha_1 \cdot \int dx [\phi_0 - \phi_F]^2 \right) \equiv \Psi_{initial} \tag{4.8}$$

with $\phi_0 \equiv \phi_F + \varepsilon^+$ and where $\alpha_2 \cong \alpha_1$. These values for the wave functionals showed up in the upper right hand side of Fig 3 and represent the decay of the false vacuum hypothesis. As mentioned this allows us to present a change in energy levels to be inversely proportional to the distance between a S-S' pair [1,3,10]

$$\alpha_2 \equiv \Delta E_{gap} \equiv \alpha \approx L^{-1} \tag{4.9}$$

We also found that in order to have a Gaussian potential in our wavefunctionals that we needed to have in both interpretations

$$\frac{(\{\ \})}{2} \equiv \Delta E_{gap} \equiv V_E(\phi_F) - V_E(\phi_T) \tag{4.10}$$

where for the Bogomol'nyi interpretation of this problem we worked with potentials (generalization of the extended Sine-Gordon model potential)[1,10]

$$V_E \cong C_1 \cdot (\phi - \phi_0)^2 - 4 \cdot C_2 \cdot \phi \cdot \phi_0 \cdot (\phi - \phi_0)^2 + C_2 \cdot (\phi^2 - \phi_0^2)^2 \tag{4.11}$$

We had a Lagrangian[15] we modified to be (due to the Bogomil'nyi inequality)

$$L_E \geq |Q| + \frac{1}{2} \cdot (\phi_0 - \phi_C)^2 \cdot \{\ \} \tag{4.12}$$

with topological charge $|Q| \to 0$ and with the Gaussian coefficient found in such a manner as to leave us with wave functionals [1,3,10] we generalized for charge density transport. This same Eq. (4.12) was more or less assumed in the Gaussian wavefunctional ansatz interpretation while we still used Eq. (4.9) and Eq. (4.10) as quasi experimental imputs into the wavefunctionals according to

$$\Psi_{i,f}\left[\phi(\mathbf{x})\right]\Big|_{\phi \equiv \phi_{ci,cf}} = c_{i,f} \cdot \exp\left\{-\int d\mathbf{x}\, \alpha \left[\phi_{Ci,f}(\mathbf{x}) - \phi_0(\mathbf{x})\right]^2\right\}, \tag{4.13}$$

In both cases, we find that the coefficient in front of the wavefunctional in Eq. (4.13) is normalized due to error function integration. This is using the pinning gap formulation of density wave transport for a S-S' pair initially pioneered by Bardeen. Furthermore, this allowed us to derive, as mentioned in another publication a stunning confirmation of the fit between the false vacuum hypothesis and data obtained for current – applied electrical field values graphs (I-E) curves of experiments initiated in the mid 1980s by Dr. John Miller, et al[13]. which lead to the modulus of the tunneling Hamiltonian being proportional to a current which we found was[1,3,10]

$$I \propto \widetilde{C}_1 \cdot \left[\cosh\left[\sqrt{\frac{2 \cdot E}{E_T \cdot c_V}} - \sqrt{\frac{E_T \cdot c_V}{E}}\right]\right] \cdot \exp\left(-\frac{E_T \cdot c_V}{E}\right) \tag{4.14}$$

This is due to evaluating our tunneling matrix Hamiltonian with the momentum version of an F.T. of the thin wall approximation, which is alluded to in Fig. 2 [1,3,10] being set by

$$\phi(k_n) = \sqrt{\frac{2}{\pi}} \cdot \frac{\sin(k_n L/2)}{k_n} \qquad (4.15)$$

This was a great improvement upon the Zenier curve fitting polynomial which was used by Miller et al[16]. We also assume a normalization of the form

$$C_i = \frac{1}{\sqrt{\int_0^{\sqrt{\frac{L^2}{2\pi}}} \exp(-2 \cdot \{\}_i \cdot \phi^2(k)) \cdot d\phi(k)}} \qquad (4.16)$$

In doing this, $\{\}_i$ refers to initial and final momentum state information of the wave functional integrands obtained by the conversion of our initial and final CDW wave functional states to a $\phi(k)$ 'momentum' basis. We evaluate for $i = 1,2$ representing the initial and final wave functional states for CDW transport via the error function

$$\int_0^{\sqrt{\frac{L^2}{2\pi}}} \Psi_i^2 \cdot d\phi(k_n) = 1 \qquad (4.17)$$

due to an error function behaving as[19]

$$\int_0^b \exp(-a \cdot x^2) dx = 1/2 \cdot \sqrt{\frac{\pi}{a}} \cdot erf(b \cdot \sqrt{a}) \qquad (4.18)$$

leading to a renormalization of the form

$$\tilde{C}_1 \equiv \frac{C_1 \cdot C_2}{2 \cdot m_{e^-}} \qquad (4.19)$$

The current expression[1,3,10,13] is a great improvement upon the phenomenological Zener current[16] expression, where $G_P$ is the limiting CDW conductance.

$$I \propto G_P \cdot (E - E_T) \cdot \exp\left(-\frac{E_T}{E}\right) \text{ if E > ET} \qquad (4.20)$$

$$0 \quad \text{otherwise}$$

Furthermore, we have that we are observing this occurring while taking into account the situation in Fig. 5 which leads to a proportionality argument we can use

The Bloch bands are tilted by an applied electric field when we have $E_{DC} \geq E_T$ leading to a S-S' pair shown in Fig. 5. The slope of the tilted band structure is given by $e^* \cdot E$ and the separation between the S-S' pair is given by, as referred to in Fig. 2 . Note that the $e^* \equiv 2 \cdot e^-$ due to the constituent components of a S-S' pair . And Fig. 2 gives us the following distance, $L$, where $\Delta_s$ is a 'vertical' distance between the two band structures tilted by an applied electric field, and $L$ is the distance between the constituent S-S' charge centers.

$$L = \left(\frac{2 \cdot \Delta_s}{e^*}\right) \cdot \frac{1}{E} \qquad (4.21)$$

So then, we have $L \propto E^{-1}$. When we consider a Zener diagram of CDW electrons with tunneling only happening when $e^* \cdot E \cdot L > \varepsilon_G$ where $e^*$ is the effective charge of each condensed electron and $\varepsilon_G$ being pinning gap energy, we find.

$$\frac{L}{x} \equiv \frac{L}{\bar{x}} \cong c_v \cdot \frac{E_T}{E} \qquad (4.22)$$

Here, $c_v$ is a proportionality factor included to accommodate the physics of a given spatial (for a CDW chain) harmonic approximation of

$$\bar{x} = \bar{x}_0 \cdot \cos(\omega \cdot t) \Leftrightarrow m_{e^-} \cdot a = -m_{e^-} \cdot \omega^2 \cdot \bar{x} = e^- \cdot E \Leftrightarrow \bar{x} = \frac{e^- \cdot E}{m_{e^-} \omega^2}$$

(4.23)

Realistically, an experimentalist[1,3,10] will have to consider that $L \gg \bar{x}$, where $\bar{x}$ an assumed reference point is an observer picks to measure where a S-S' pair is on an assumed one-dimensional chain of impurity sites.

## V. CONCLUSION: SETTING UP THE FRAMEWORK FOR A FIELD THEORETICAL TREATMENT OF TUNNELING

We have, in the above document identified pertinent issues needed to be addressed in an analytical treatment of Charge Density Wave transport. First, we should try to have a formulation of the problem of tunneling which has some congruence with respect to the 'False Vacuum' hypothesis of Sidney Coleman[9]. We make this statement based upon the abrupt transitions made in a multi-chain model of Charge Density Wave tunneling which are in form identical to what we would expect in a thin wall approximation of a boundary between true and false vacuums.

Prior researchers/authors have given very reasonable attempts to analyze density wave transport from a field theoretic standpoint. Kazumi Makis excellent start in 1977[8] was marred though because he did not have experimental data present to Miller and other researchers later on about the importance of a threshold field $E_T$ for initiation of density wave nucleation and he did not include it explicitly in his calculations. We should note that several quantum tunneling approaches to this issue have been proposed. One [4] is to use functional integrals to compute the Euclidean action ("bounce") in imaginary time. This permits one to invert the potential and to modify what was previously a potential barrier separating the false and true vacuums into a potential well in Euclidean space and imaginary time. The decay of the false vacuum is a potent paradigm for describing decay of a metastable state to one of lower potential energy. In condensed matter, this decay of the false vacuum method has been used[20] to describe nucleation of cigar-shaped regions of true vacuum with soliton-like domain walls at the boundaries in a charge density wave .We use the Euclidian action so that we may invert the potential in order to use WKB semi-classical procedures for solving our problem. Another approach[21], using the Schwinger proper time method, has been applied by other researchers to calculate the rates of particle-antiparticle pair creation in an electric field[22] for the purpose of simplifying transport problems. What we are proposing here is a synthesis of several methods, plus additional insight as to the topological charge dynamics of density wave transport which were neglected in prior attempts to analyze this problem fully.

We explicitly argue that a tunneling Hamiltonian based upon functional integral methods is essential for satisfying necessary conditions for the formation of a S-S' pair. The Bogomil'nyi inequality stresses the importance of the relative unimportance of the driving force $E_2 \cdot (\phi_n - \Theta)^2$, which we drop out in our formation of a soliton (anti soliton) in our multi-chain calculation. In addition, we argue those normalization procedures, plus assuming a net average value of the $\Delta'(1 - \cos[\phi_n - \phi_{n-1}]) \to \frac{\Delta'}{2} \cdot [\phi_n - \phi_{n-1}]^2 +$ small terms as seen in our analysis of the contribution to the Peierls gap contribution to S-S' pair formation in our Gaussian $\psi \propto c \cdot \exp(-\beta \cdot \int L \, dx)$ representation of how S-S' pairs evolve in a pinning gap transport problem for charge density wave dynamics. The overall convergence of a numerical scheme to represent multi-chain contributions to the analysis of this problem, gives a Josephon junction flavor to our analysis. It also underlies the formation of solitons (anti solitons) which was used by us as the underpinnings of the S-S' pairs used to give more detailed structure to the field theoretic analysis of this important problem.. This work in itself is a step forward from the initially classical analysis offered by Gruner[23] Furthermore, what is done here is a simpler treatment of transport modeling as is seen in older treatment in the literature[24] and also makes full use of Bardeens[6] pinning gap arguments, which is a more direct analysis of density wave dynamics than the typical CDW literature presented earlier. Also it improves upon the simple minded current calculations done in the literature[25] based upon simplistic quantum measurement calculations .We would like to in future work to examine the implications of Sidney Colemans[25] references to not needing a renormalization other than that needed for zero point energy in his paper 'More about the Massive Schwinger model'[26], but we do not think this will affect the *I-E* plots derived analytically and referenced in this publication.

## VI. ACKNOWLEGEMENTS

The author wishes to thank Dr. John Miller for introducing this problem to him in 2000, as well as for his discussions with regards to the role bosonic states play in affecting the relative power law contribution of the magnitude of the absolute value of the tunneling matrix elements used in the current calculations. In addition, Dr. Leiming Xie highlighted the importance of Eq. (4.14) as an improvement over Eq.(4.20) in this problems evaluation. Furthermore, as Dr. Xie noted, the fact that the prior Zener curve yielded negative values for current values as the electric field was below a threshold value, while Eq.(4.14) above has no such pathology is extremely illuminating physics which deserves further investigation.

# APPENDIX I: ADDITIONAL COMPUTER SIMULATION MATERIAL W.R.T. MULTI-CHAIN CDW TRANSPORT & THE LARGE TIME SCALE RESONANCE BEHAVIOR OFA SINGLE CDW CHAIN

In our discussion about the single-chain simulation material, we looked at a second numerical scheme[3]. the Dunford-Frankel and 'fully implicit' allows us to expand the time step even further. Then, the 'massive Schwinger model' equation[3,6,26] has[3]:

$$\phi(j,n+1) = \frac{2 \cdot \widetilde{R}}{1 + 2 \cdot \widetilde{R}} \cdot (\phi(j-1,n) - \phi(j+1,n)) + \frac{1 - 2 \cdot \widetilde{R}}{1 + 2 \cdot \widetilde{R}} \cdot \phi(j,n-1) - i \cdot \Delta t \frac{V(j,n)}{\hbar} \phi(j,n) \tag{1}$$

where one has $\widetilde{R} = -i \cdot \Delta t \frac{\hbar}{2 \cdot D \cdot (\Delta x)^2}$ .The advantage of this model is that it is second order accurate ,explicit ,and unconditionally stable, so as to avoid numerical blow up behavior .One then gets resonance phenomena as represented by Fig 1 and Fig 6.This is to put it mildly quite unphysical and necessitates making changes, which we will be presenting in this manuscript

This failure necessitated going to multi-chain simulations .Now, our Peierls gap energy term $\Delta^{'}$ [7] was added to the massive Schwinger equation model[2,6,26] precisely due to the prior resonance behavior with a one chain computer simulation. We can now look at the situation with more than one chain. To do so, take a look at a Hamiltonian with Peierls condensation energy used to couple adjacent chains (or transverse wave vectors):

$$H = \sum_n \left[ \frac{\Pi_n^2}{2 \cdot D_1} + E_1[1 - \cos\phi_n] + E_2(\phi_n - \Theta)^2 + \Delta[1 - \cos(\phi_n - \phi_{n-1})] \right] \tag{2}$$

and

$$\Pi_n = (h/i) \cdot \partial/\partial\phi_n \tag{2a}$$

and when we will use wave functions which are

$$\Psi = N \cdot \prod_j \left( a_1 \exp(-\alpha \cdot \phi_j^2) + a_2 \exp(-\alpha(\phi_j - 2 \cdot \pi)^2) \right) \tag{3}$$

with a two-chain analogue of [3]

$$\Psi_{two\,chains} = N \cdot \prod_{n=1}^{2} \left( a_1 \exp(-\alpha \cdot \phi_j^2) + a_2 \exp(-\alpha(\phi_j - 2 \cdot \pi)^2) \right) \tag{4}$$

If so, we put in the requirement of quantum degrees of freedom so that one has for each chain for a two dimensional case[3]

$$|a_1|^2 + |a_2|^2 = 1 \tag{5}$$

which provides coupling between 'nearest neighbor' chains. In doing so, we are changing the background potential of this problem from a situation given in Fig. 6, to a different situation where one has multiple soliton pairs that are due to the $\Delta'$ term in which has two double well band structures given which permit the existence of tunneling due to the double well. band structure[3] We also have that $\alpha \approx \frac{1}{\sqrt{soliton\ width}}$. For 'phase co-ordinate' $\phi_j$, $\exp(-\alpha \cdot \phi_j^2)$ is an un renormalized Gaussian representing a 'soliton' (anti soliton) centered at $\phi_j = 0$, and a probability of being centered there [3] given by $|a_1|^2$. Similarly, $\exp(-\alpha \cdot (\phi_j - 2 \cdot \pi)^2)$ is an un renormalized Gaussian representing a 'soliton'(anti soliton) centered at $\phi_j = 2 \cdot \pi$ with a probability of occurrence[3] at this position given by $|a_2|^2$. We can use Eq. 5 of this appendix to represent the total probability that one has some sort of tunneling through a potential given by Eq. 2 of this appendix with the potential dominated by the term $\Delta'$ which dominates the dynamics we will see numerically in the following simulations given below.

One then can draw, with the help of a 'minimized' energy 'functional' when we generalize Eq. 2 to have a potential energy cusp with the generalized two chain energy in the form of Eq. 6, in a double potential energy well band structure plot showing up in my dissertation. This used[3]

$$E(\Theta) = \langle \Psi_{two\,chains} | H_{two\,chains} | \Psi_{two\,chains} \rangle \tag{6}$$

This is, in form, substantially the same diagram given by Miller, et al[2]. The importance of Eq. (6) is that it appears one needs the term $\Delta'$ given in Eq. (4) in order to get this band structure. The situation done with a simulation with $\Delta'(1-\cos[\phi_2 - \phi_1])^3$ included is, with Fortran 90, complicated since this would ordinarily imply coupled differential equations, which are extremely unreliable to solve numerically. For a number of reasons, one encounters horrendous round off errors with coupled differential equations solved numerically in Fortran. So, then the problem was done, instead, using Mathematica software which appears to avoid the truncation errors Fortran 90 presents us if we use a p.c. with standard techniques. Here is how the problem was presented before being coded for Mathematica: where one has $E_1 = E_p = $ pinning energy, $E_2 = E_c = $ charging energy, and $\Delta' \cdot [1 - \cos(\phi_2 - \phi_1)]$ represents coupling between "degrees of freedom" of the two chains. The wave function used was set to a different value than given in Eq. (4)

$$\Psi_m(\phi_i) = \sum_{m=-2}^{2} b_m \exp(-\alpha(\phi_i - 2 \cdot \pi \cdot m)) \tag{7}$$

with

$$\sum_{m=-2}^{2} b_m^2 = 1 \tag{8}$$

we obtained a minimum energy 'band structure' with five adjacent parabolic arcs. We obtain a 'minimum' energy out of this we can write as

$$E = E_{min} = \langle \Psi | \hat{H} | \Psi \rangle \tag{9}$$

where $D_1 = 174.091$, $E_p = .00001$, $E_c = .000001$ and $\Delta' = .005$ for Hamiltonian

$$\hat{H}_{two\,chains} = \sum_{n=1}^{2} \left[ \frac{\Pi_n^2}{2 \cdot D_1} + E_1[1 - \cos\phi_n] + E_2(\phi_n - \Theta)^2 + \Delta' \cdot [1 - \cos(\phi_n - \phi_{n-1})] \right] \tag{10}$$

where minimum energy curves are set by the coefficients of the two wave functions, which are set as $b_{-2}, b_{-1}, b_0, b_1, b_2; c_{-2}, c_{-1}, c_0, c_1, c_2; \alpha$ (which happens to be the wave parameter for Eq. (10). This leads to an energy curve given in Fig. 7 where there are five, not two local minimum values of the energy as given in the plot given initially in my dissertation. It is a reasonable guess that for additional chains (i.e. if m bracketed by numbers > 2) that the number of local minimum values will go up, provided that one uses a modified version of numerical simulation wave function probability as given in Eq. (8) for Eq. (7) of this appendix. We did the following to plot an average <phi> value, which we will represent in Eq. (12) below. The easiest way to put in a time dependence in the Hamiltonian Eq. (10) is to provisionally set $\epsilon = a_D t$ for the graphics presented, $a_D = 0.67\,\text{M Hz}$

If we set $\Psi \equiv \Psi(\phi_1, \phi_2, \epsilon)$ which has an input from the Hamiltonian $\hat{H}_{two\,chains}$ then we can set up an average phase, which we will call

$$\Phi = \frac{1}{2}(\phi_1 + \phi_2) \qquad (11)$$

where we calculate a mean value of phase given by [3,7]

$$\langle \Phi(\Theta) \rangle = \int_{-\eta\pi}^{\eta\pi} \int_{-\eta\pi}^{\eta\pi} d\phi_1 d\phi_2 \, \frac{1}{2} \cdot (\phi_1 + \phi_2) |\Psi(\phi_1, \phi_2, \Theta)|^2 \qquad (12)$$

The integral $\langle \Phi(\Theta) \rangle$ was evaluated by 'Nintegrate' of Mathematica, and was graphed against $\epsilon$ in Fig. 8, with $\eta = 20$. These total sets of graphs put together are strongly suggestive of tunneling when one has $\Delta \neq 0$ in $\hat{H}_{two\,chains}$.

The simulation results of Fig. 9 are akin to a thin wall approximation leading to a specific shape of the soliton – anti soliton pair in 'phase' space which is also akin to when we have abrupt but finite transitions after long periods of stability.[1,2]

# Figure Captions

Fig 1: Beginning of resonance phenomena due to using the traditional Crank-Nickelson numerical iteration scheme of the one chain model. Phi refers to a time dependent phase value due to a single chain approximation.

Fig 2: The above figures represent the formation of soliton-anti soliton pairs along a 'chain'. The evolution of phase is spatially given by
$\varphi(x) = \pi \cdot [\tanh b(x - x_a) + \tanh b(x_b - x)]$

Fig 3: Evolution from an initial state $\varphi_i$ to a final state $\varphi_f$ for a double-well potential (inset) in a quasi 1-D model, showing a kink-anti kink pair bounding the nucleated bubble of true vacuum. The shading illustrates quantum fluctuations about the optimum configurations of the field $\varphi_F$ and $\varphi_T$, while $\varphi_0(x)$ represents an intermediate field configuration inside the tunnel barrier. This also shows the direct influence of the Bogomil'nyi inequality in giving a linkage between the 'distance' between constituents of a 'nucleated pair' of $S$-$S'$ and the $\Delta E$ difference in energy values between $V(\varphi_F)$ and $V(\varphi_T)$ which allowed us to have a 'Gaussian' representation of evolving nucleated states.

Fig 4: Experimental and theoretical predictions of current values versus applied electric field. The dots represent a Zenier curve fitting polynomial, whereas the blue circles are for the $S$-$S'$ transport expression derived with a field theoretic version of a tunneling Hamiltonian. This explains earlier data collected by Miller, Tucker, et al. Also, the classical current gives a negative value for applied electric fields below $E_T$

Fig 5    This is a representation of 'Zener' tunneling through pinning gap with band structure tilted by applied E field

Fig 6:   Figure presented completes proof that one chain does not permit tunneling, using Dunford- Frankel numerical scheme for large time stepping.

Fig 7:   Determining band structure via a Mathematica 8 program, and with wave functions given by Eq. (7) of Appendix I

Fig 8:   Phase vs. Є ,according to the predictions of the 'multi- chain'-tunneling tunneling model.

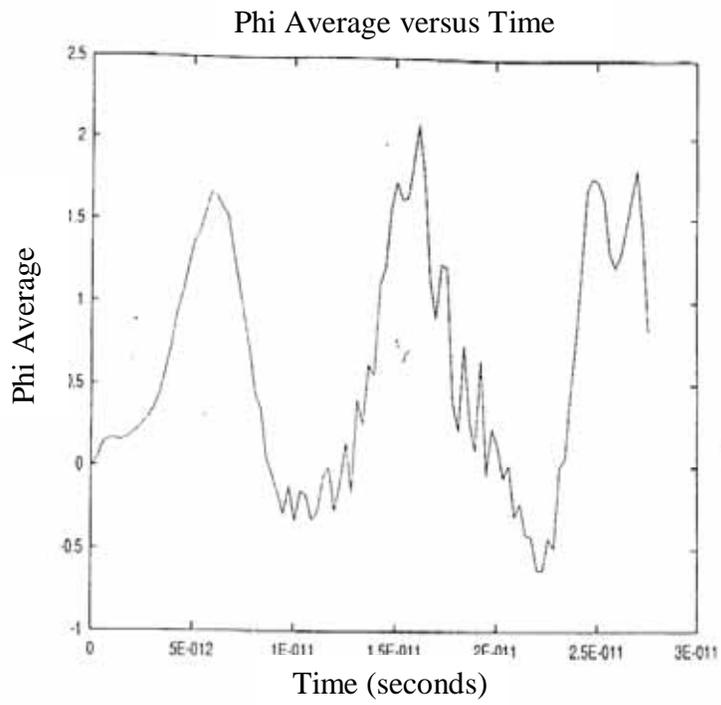

**Figure 1
Beckwith**

# CDW and its Solitons

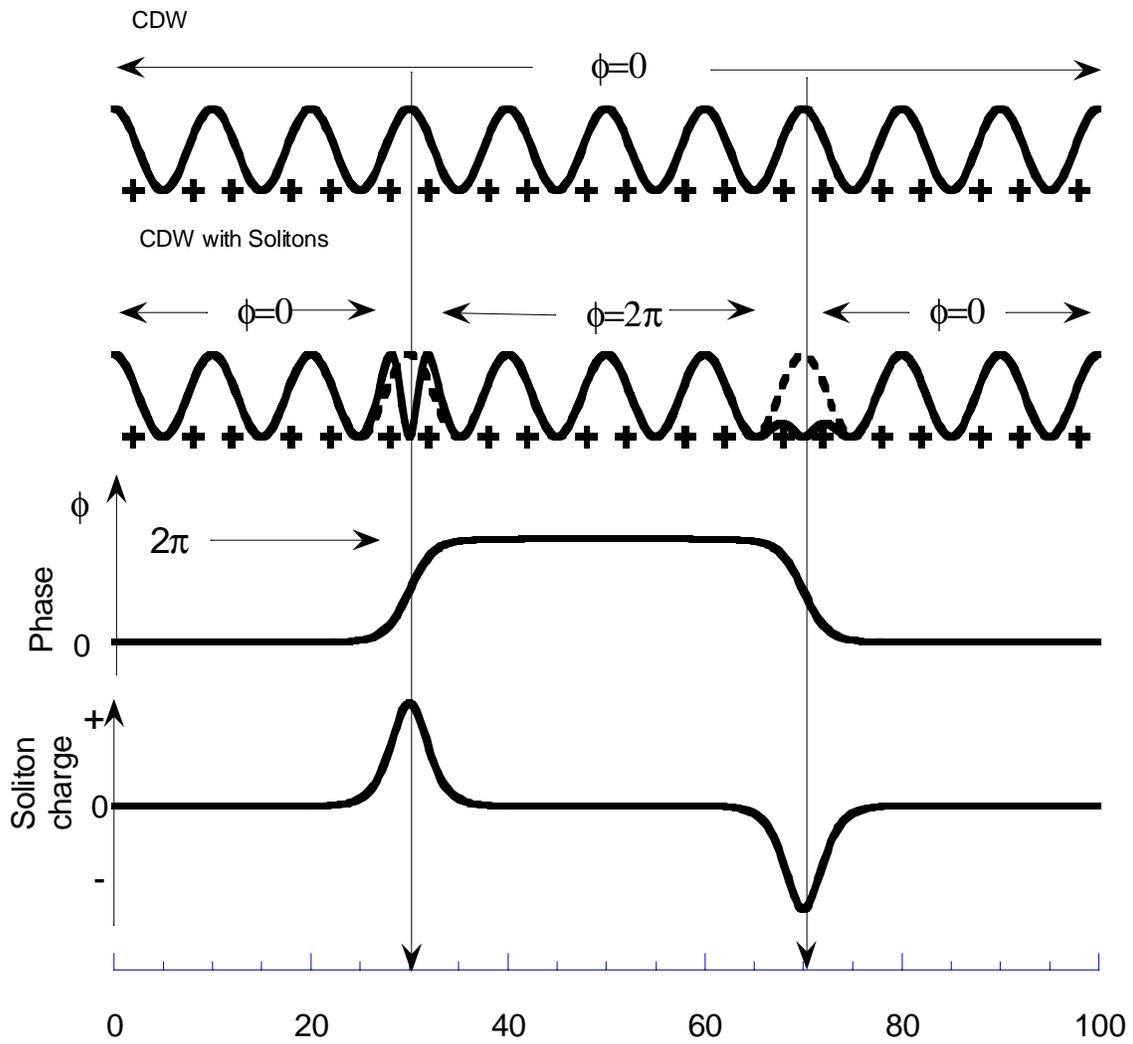

Figure 2
Beckwith

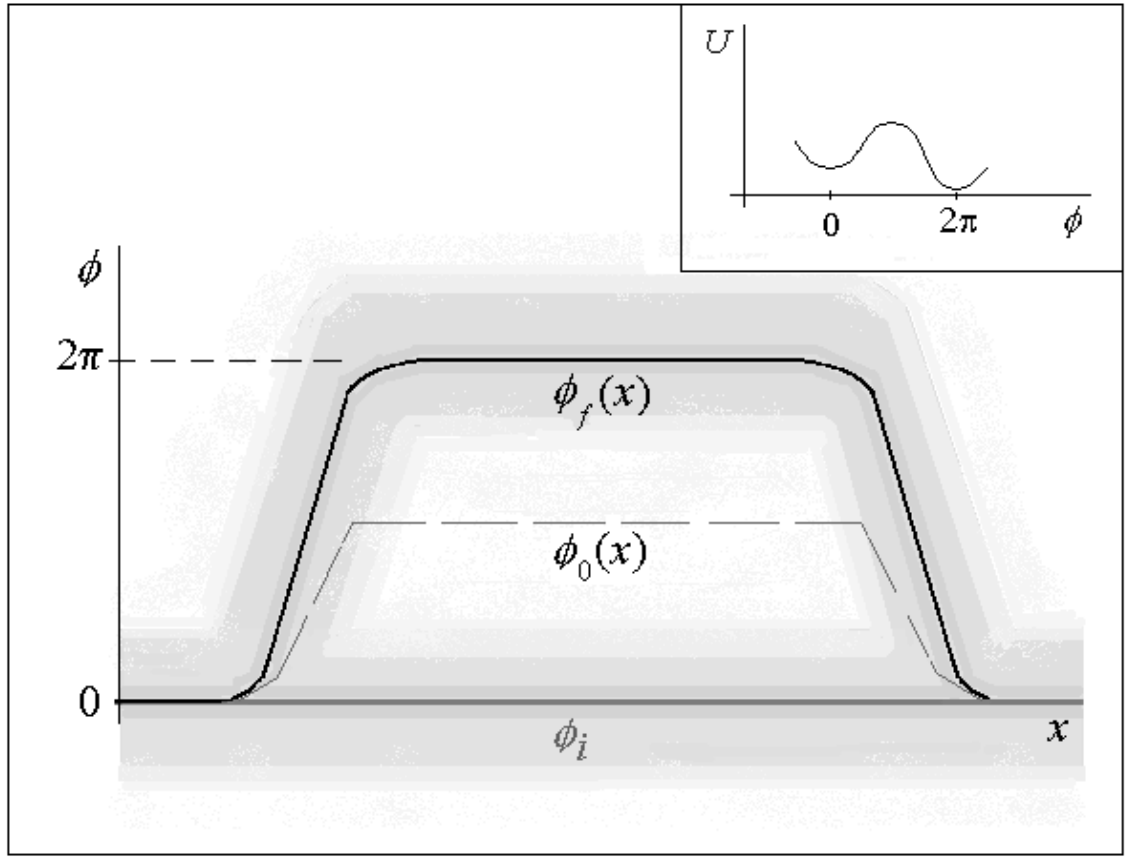

**Figure 3
Beckwith**

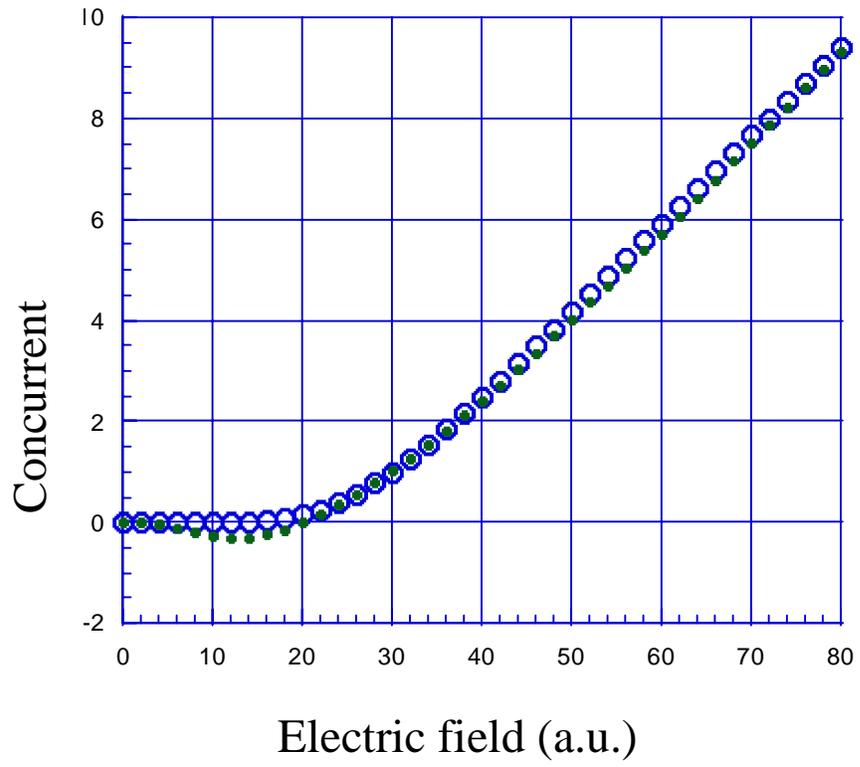

**Figure 4
Beckwith**

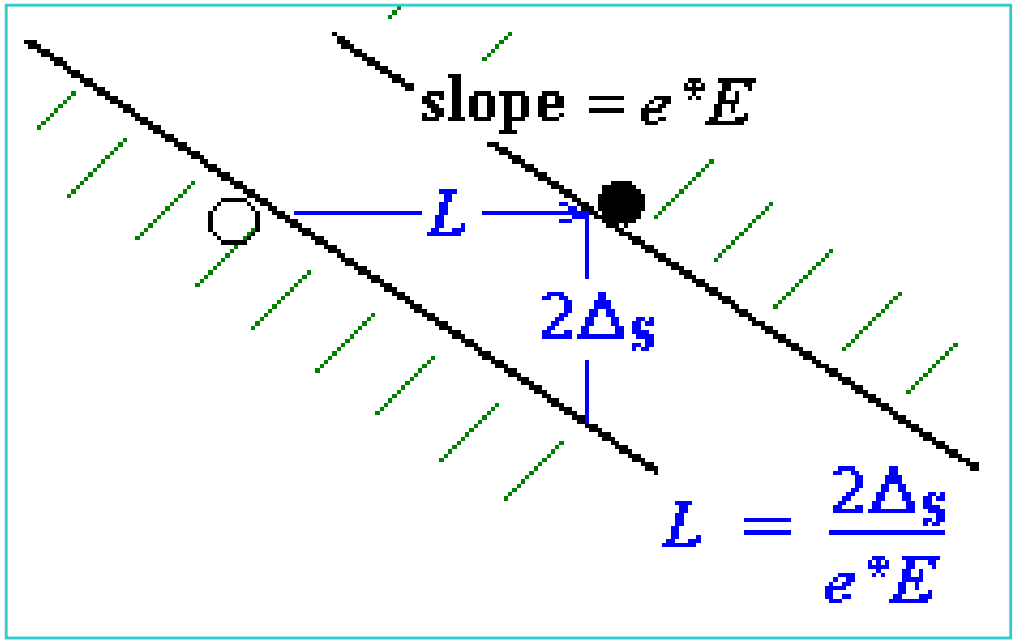

**Figure 5
Beckwith**

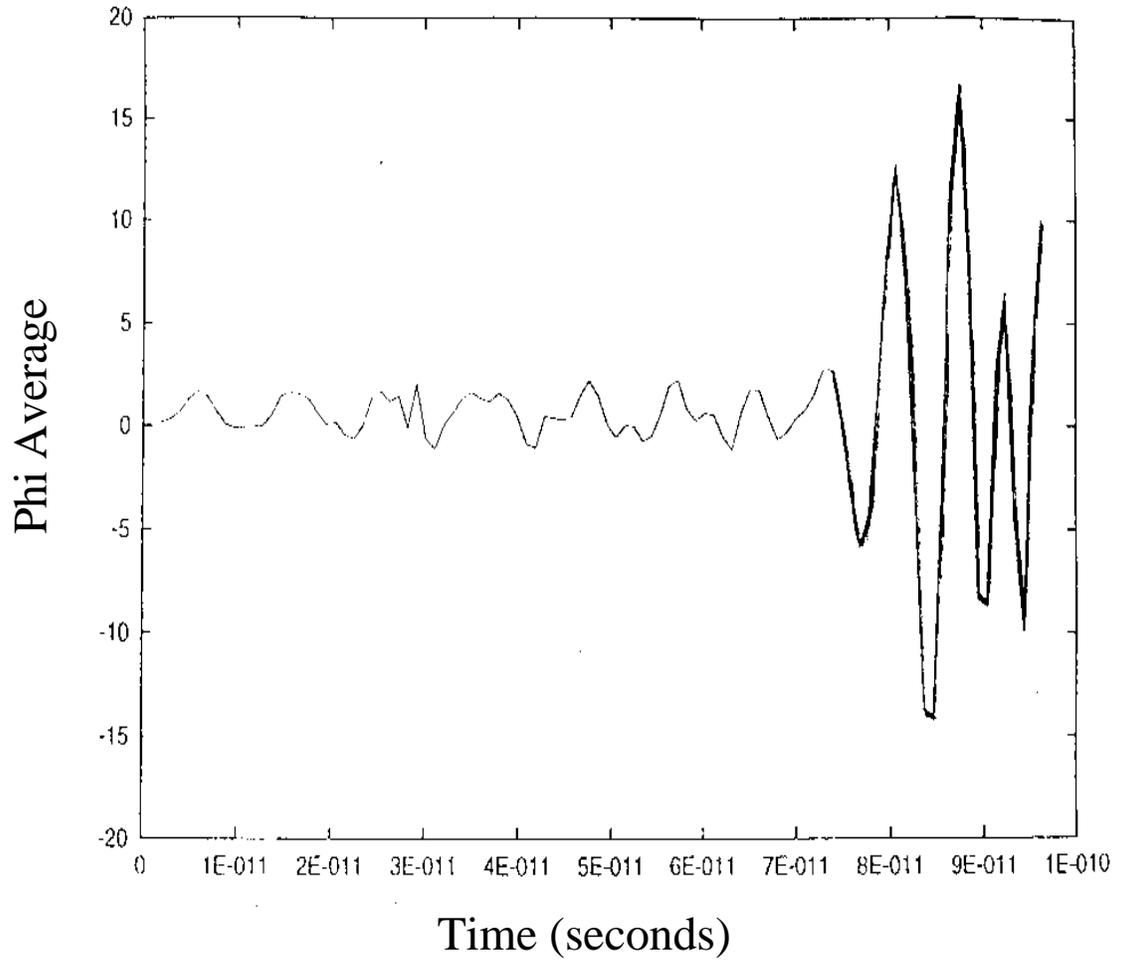

**Figure 6
Beckwith**

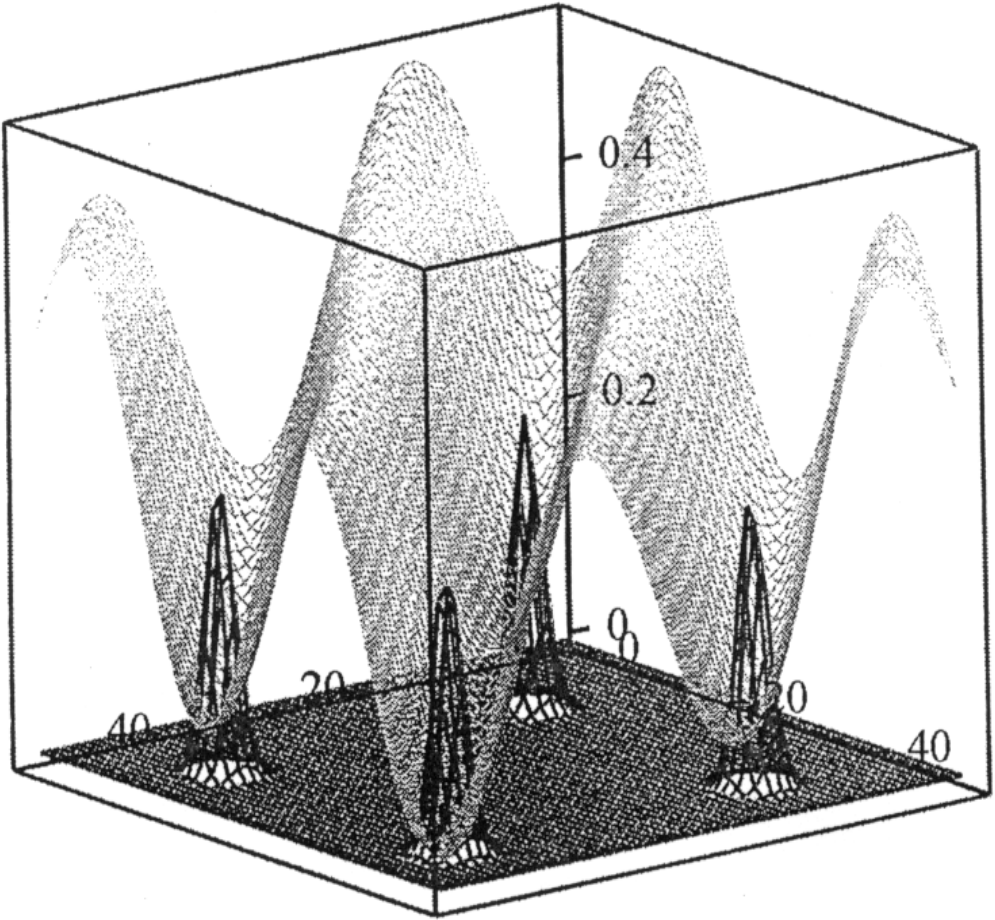

**Figure 7
Beckwith**

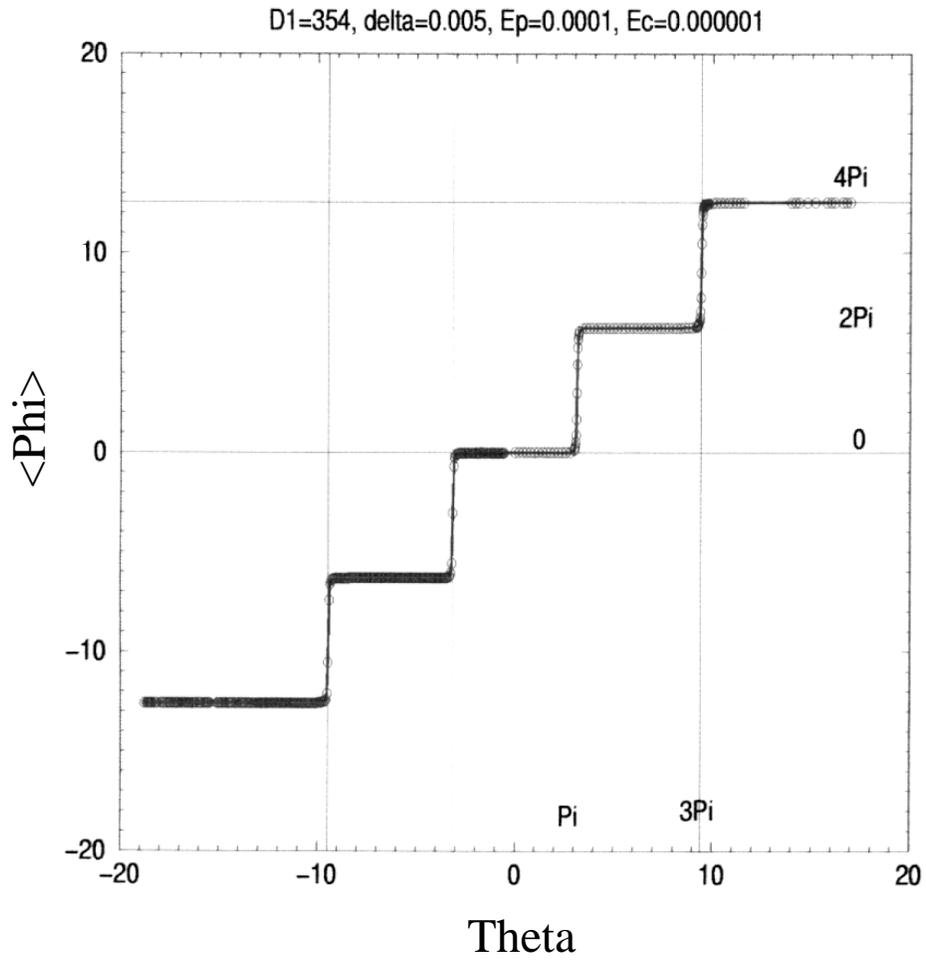

**Figure 6
Beckwith**